\documentclass[preprint,12pt]{elsarticle}

\usepackage{amssymb}
\usepackage{siunitx}
\usepackage{soul}
\usepackage{rotating}
\usepackage{amsmath}

\RequirePackage{xurl}

\journal{Progress in Photovoltaics}

\begin{document}
\emergencystretch 3em
\begin{frontmatter}



\title{Energy Yield and Lifetime Climate Classification via Machine Learning for Optimizing Photovoltaic Module Design and Materials}

\author[TUDelft]{Youri Blom\corref{cor1}}
\author[UniversityNaples]{Sofia Dutto}
\author[TUDelft]{Alexandru Costache}
\author[TUDelft]{Rowan Richie}
\author[TUDelft]{Ruben Pelsser}
\author[TUDelft]{Wesley Berger}
\author[TUDelft]{Jing Sun}
\author[TUDelft]{Rudi Santbergen}
\author[TUDelft]{Olindo Isabella}
\author[TUDelft]{Malte Ruben Vogt\corref{cor2}}

\affiliation[TUDelft]{organization={Delft University of Technology},
            addressline={Mekelweg 4},
            city={Delft},
            postcode={2628CD},
            country={The Netherlands}}
\affiliation[UniversityNaples]{organization={University of Naples Federico II},
            addressline={Via Claudio, 21},
            city={Naples},
            postcode={80125},
            country={Italy}}

\cortext[cor1]{Y.Blom@tudelft.nl}
\cortext[cor2]{M.R.Vogt@tudelft.nl}

\begin{abstract}
To resiliently and sustainably meet our future energy demand, photovoltaic (PV) modules must be deployed across a broad and diverse range of geographical regions with varying operating conditions. 
As these conditions strongly affect both performance and optimal system design, a dedicated PV-specific climate classification can be of great use.
In this work, we develop a climate classification framework tailored to PV applications using a variety of machine learning (ML) techniques. 
Building on previous studies, our approach incorporates both energy yield, and for the first time, also the module lifetime with climate dependent degradation.
We generate an interpolated dataset containing twelve input features and two target variables (i.e. energy yield and module lifetime). 
Feature importance analysis shows that annual global horizontal irradiation and ambient temperature are the most influential predictors. 
The most accurate regression model achieves root mean square errors (RMSE) of 0.007~\unit{\mega\watt\hour} for energy yield and 1.5 years for lifetime prediction.
The calculated feature importance scores are then integrated into a hierarchical clustering framework, resulting in 6 primary climate clusters—Tropical, Desert, Continental, Temperate, Boreal, and Polar—and 15 corresponding subclusters.
Our analysis shows that the low‑temperature continental climate offers the highest discounted lifetime energy yield. 
These results can support a wide range of applications, including PV module optimization, system siting decisions, and comparative performance studies.
\end{abstract}

\begin{graphicalabstract}
\includegraphics{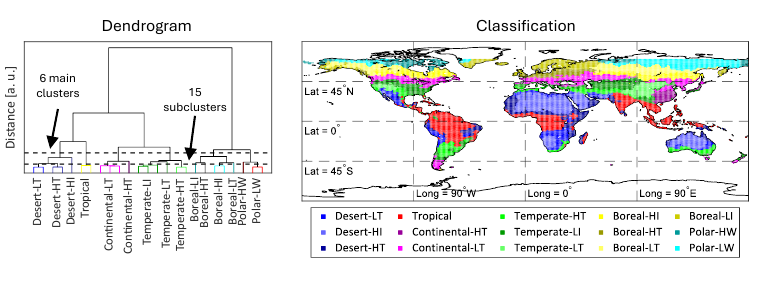}
\end{graphicalabstract}

\begin{highlights}
\item A climate classification is created covering both energy yield and lifetime aspects
\item Annual GHI and mean ambient temperature are the most important features
\item Six main clusters and fifteen subclusters are found to be optimal
\item The low-temperature continental climate has the highest lifetime energy yield
\end{highlights}


\end{frontmatter}



\section{Introduction}\label{sec:Introduction}
To meet future energy demand in a resilient and sustainable way, the global photovoltaic (PV) capacity must grow at an annual rate of 25\% over the next decade~\cite{Haegel2023PhotovoltaicsOption}. As a result, PV modules will need to be deployed across an increasingly diverse set of geographical regions, each characterized by widely varying operating conditions.
These varying operating conditions strongly influence the performance and behavior of PV modules, affecting both their operating efficiency~\cite{Rahman2015EffectsEfficiency, Hasan2022EffectsReview} and degradation rates~\cite{Barretta2025EffectClimates, Jordan2016CompendiumRates}. 
Consequently, PV manufacturers have already started customizing the bill of materials (BoM) for modules intended for different environmental conditions~\cite{TaiyangNewsShielModules, PVmagazineModulesAboveSeaLevel}.
It has also been demonstrated that the optimal cell design—such as wafer thickness in crystalline silicon (c-Si) cells~\cite{Ziar2024AMarkets} or the bandgap energy of the perovskite top cell in perovskite/silicon tandems~\cite{Blom2025OptimizationModule}—varies across locations and climate types. 
Similarly, the International Energy Agency PV Power Systems program has published specific guidelines for operation and maintenance practices tailored to different climate zones~\cite{IEA2022_GuidlinesClimates}.

To simplify the task of customizing solar cell and BoM designs for specific locations, a climate classification can be developed that groups regions with similar operating conditions. 
Such a classification would reduce the problem from potentially designing a PV module for each location to only a limited number of module types, each optimized for a particular climate zone.
Several PV-specific climate classifications have already been proposed in the literature. 
Dash~\textit{et al.}~\cite{Dash2017ATechnologies} divided India into 7 climatic zones based on annual daytime irradiance and ambient temperature. 
The most widely used global climate classification is the K$\ddot{\textrm{o}}$ppen--Geiger (KG) system, originally proposed by Kottek~\textit{et al.}~\cite{Kottek2006WorldUpdated}, which divides the world into 6 main climate types and 14 sub‑climates.
Because the KG classification is based solely on ambient temperature and precipitation, several studies have extended it to include solar irradiance. 
Ascencio-Vásquez~\textit{et al.}~\cite{Ascencio-Vasquez2019MethodologyPerformance} developed the K$\ddot{\textrm{o}}$ppen--Geiger Photovoltaic (KGPV) classification, resulting in 12 PV‑relevant climate zones. 
Similarly, Skandalos~\textit{et al.}~\cite{SKSkandalos2022BuildingIncreases} adapted the KG system for building‑integrated PV applications.

Although these extended climate classifications represent meaningful improvements over the original KG system, several parameters relevant to PV performance—such as wind speed and ultraviolet irradiance—are still not included. 
Moreover, because solar energy was not the focus of the original KG classification, the derived PV-specific adaptations may not be fully optimal.
To address these limitations, Triana de las Heras~\textit{et al.}~\cite{TrianadelasHeras2025AClassification} developed a new PV climate classification map using a range of Machine Learning (ML) techniques. 
The use of ML in PV-related research has increased rapidly in recent decades~\cite{Alcaniz2023TrendsLearning, Trina2021APV}, largely due to its strong capability for handling large datasets and identifying complex patterns~\cite{Dahiya2022ATechniques}.
In those works, a regression-based ML model is first employed to determine the most influential environmental variables (features) for predicting energy yield at global scale.
These key features are then used in a clustering algorithm to group locations with similar characteristics.
While this approach is promising, the study relies on relatively simple ML methods—such as linear regression and k-means clustering. 
In addition, the resulting climate classification focuses solely on energy yield and does not incorporate lifetime performance or degradation behavior.

This work builds on the approach of Triana de las Heras~\textit{et al.} by developing a ML-based climate classification that focuses on both the energy yield and lifetime performance of c-Si modules. 
By combining ML methods of varying complexity, the most relevant environmental parameters are identified and subsequently used to cluster locations with similar characteristics.
The paper is structured as follows: Section~\ref{sec:Dataset} describes the dataset used to construct the climate classification; Section~\ref{sec:Methodology} outlines the methodology of this study, covering both the physical models employed to generate the dataset and the ML techniques used in the analysis; Section~\ref{sec:Results} presents the obtained results, beginning with the accuracy of the ML methods and the resulting feature importance, followed by the final clustering outcomes; lastly, Section~\ref{sec:Conclusion} summarizes the main findings and offers concluding remarks.

\section{Dataset}\label{sec:Dataset}
To develop the climate classification, a global dataset is required that includes both environmental conditions and PV system performance across the world.
The Meteonorm database~\cite{remund2020meteonorm} is used to extract hourly environmental data from all available weather stations worldwide.
Based on this data, 12 environmental characteristics (features) and two performance metrics (targets) are defined.
Note that this database is used instead of the available dataset from Triana de las Heras~\textit{et al.}~\cite{TrianadelasHeras2025AClassification}, as hourly data is needed for an accurate calculation of both targets.
Because the initial sampling of locations is not uniformly distributed across the Earth, spatial interpolation is applied to correct for geographical imbalances. 
Finally, the correlations between all features and targets are evaluated.

\subsection{Features}
To describe the hourly environmental conditions, features are defined that are potentially relevant to both energy yield and lifetime performance. 
Table~\ref{tab:Features} lists all features considered in this study. 
The selection is primarily based on the features used by Triana de las Heras~\textit{et al.}~\cite{TrianadelasHeras2025AClassification}, supplemented with additional features that have a physical relationship with the chosen targets.
The mathematical definitions of these features, as well as their global spatial distributions, are provided in~\ref{App:FeatureDefinitions}. 
All features are normalized to have zero mean and unit standard deviation to facilitate a better comparison.

\begin{table}[ht]
    \centering
    \caption{The features that are considered for the climate classification.}
    \begin{tabular}{p{4cm}| p{8cm}}
    \textbf{Feature} & \textbf{Description}  \\
    \hline
    $\textrm{T}_{\textrm{mean}}$[\unit{\celsius}] & Annual mean ambient temperature\\
    $\textrm{T}_{\textrm{min}}$[\unit{\celsius}] & Mean ambient temperature of coldest month\\
    $\textrm{T}_{\textrm{max}}$[\unit{\celsius}] & Mean ambient temperature of warmest month\\
    $\textrm{T}_{\textrm{diff}}$[\unit{\celsius}] & Mean value of daily temperature variation\\
    $\textrm{GHI}_{\textrm{ann}}$[\unit{\mega\watt\hour\per\meter\squared}] & Annual received Global Horizontal Irradiance (GHI)\\
    $\textrm{GHI}_{\textrm{min}}$[\unit{\kilo\watt\hour\per\meter\squared}] & Minimum monthly received GHI\\
    $\textrm{GHI}_{\textrm{max}}$[\unit{\kilo\watt\hour\per\meter\squared}] & Maximum monthly received GHI\\
    $\textrm{DNI}_{\textrm{share}}$[\%] & Share of direct irradiance\\
    $\textrm{WS}_{\textrm{mean}}$[\unit{\meter\per\second}] & Annual mean wind speed\\
    $\textrm{AM}_{\textrm{eff}}$[-] & Effective air mass\\
    $\textrm{RH}_{\textrm{mean}}$[\%] & Annual mean relative humidity\\
    $\textrm{UV}_{\textrm{ann}}$[\unit{\kilo\watt\hour\per\meter\squared}] & Annual received ultraviolet (UV) irradiation\\
    \end{tabular}
    \label{tab:Features}
\end{table}

\subsection{Model targets}
The two targets used to evaluate PV module performance are the annual energy yield and module lifetime.
Since, to the best of the authors' knowledge, no global dataset of measured energy yield and module lifetime exists, physical models are used to generate these quantities.
We employ the PVMD Toolbox~\cite{Vogt2022IntroducingSystems, Blom2026ImprovingSystems} to simulate the energy yield and lifetime of c-Si modules installed all around the world.

The PVMD Toolbox is a modeling framework for simulating the outdoor performance of PV systems.
It consists of sequential steps that simulate different aspects of the PV system, resulting in the hourly electricity production~\cite{Blom2026ImprovingSystems}.
In recent work~\cite{Blom2026CombiningModules}, the PVMD Toolbox has been extended for degradation analysis, allowing for the calculation of the module lifetime.
Here, module lifetime is defined as the time at which the module performance decreases to 80\% of its initial performance due to degradation effects.
As some values for the module lifetime exceeds the commonly expected range of 30-40 years~\cite{VDMA2025_ITRPV}, we have performed a validation in~\ref{App:LifetimeValiation}.
This shows that the simulated degradation rates are consistent with field measurements.

These physical models are used to calculate the annual energy yield and module lifetime of c-Si modules at different locations across the world, as shown in Figure~\ref{fig:Targets}.
The module layout and the Standard Test Conditions (STC) characteristics of the considered PV module can be found in the~\ref{App:ConsideredModule}.
To accelerate the computational time of these simulations, the calculations are carried out on the Delft High Performance Computing Centre~\cite{DHPC2024}.


\subsection{Spatial interpolation}
\begin{sidewaysfigure}[p]
    \centering
    \includegraphics[width=\linewidth]{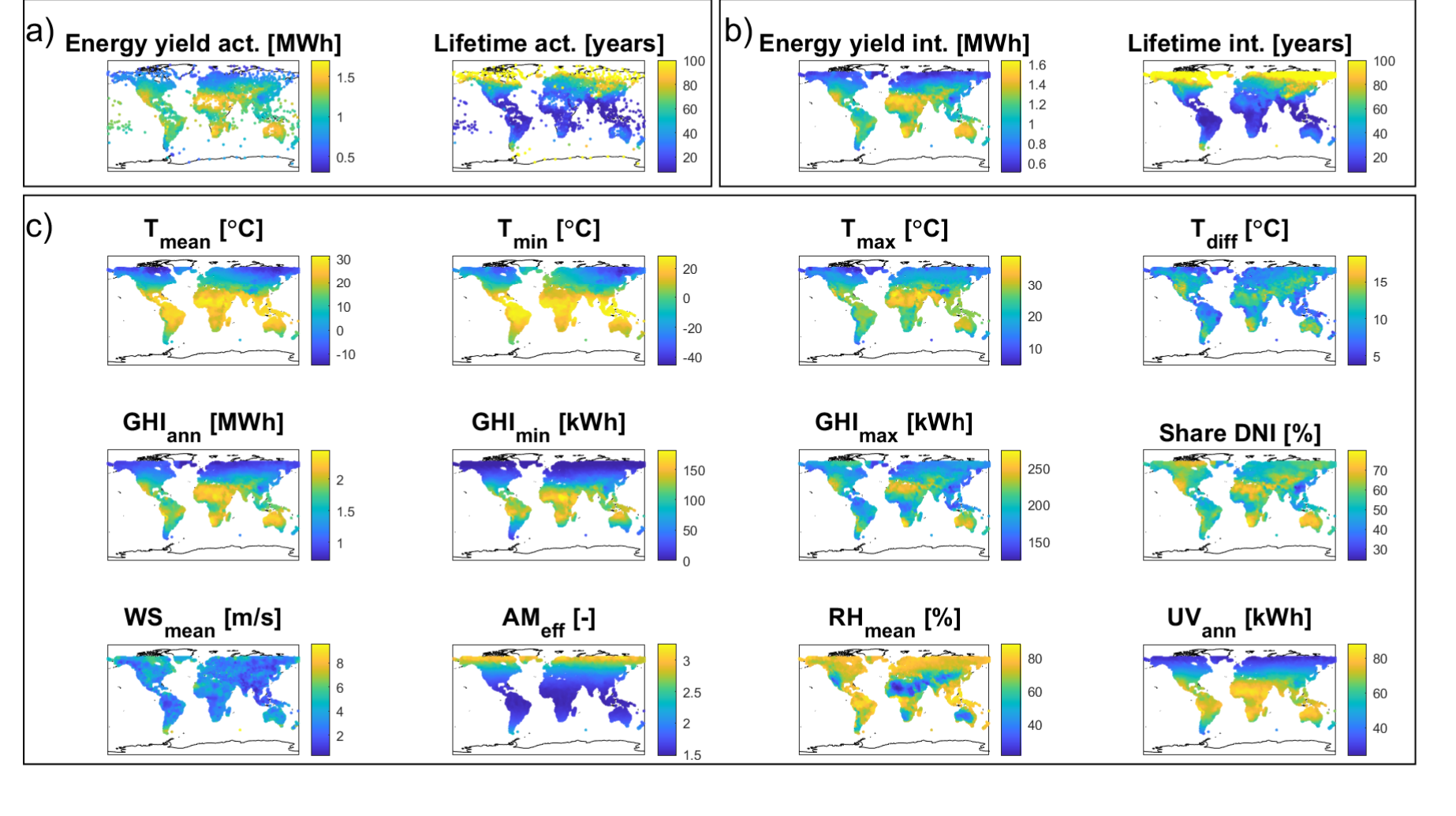}
    \caption{a) The geographical distribution of the simulated energy yield and module lifetime at the locations of the available weather stations. b) The interpolated values of energy yield and module lifetime for all coordinates. c) The interpolated values for all features.}
    \label{fig:Targets}
\end{sidewaysfigure}

As shown in Figure~\ref{fig:Targets}a, the simulated locations are unevenly distributed across the globe. 
This non-uniformity may introduce bias into the climate classification, as regions with a higher density of data points could disproportionately influence the results. 
To mitigate this effect, a nearest-neighbor interpolation approach~\cite{Shepard1968AData} (using three neighbors) is applied to the targets and all features, resulting in a more spatially uniform dataset.
Figure~\ref{fig:Targets} presents the geographical datasets after spatial interpolation for the targets. 
Both the original and interpolated datasets are provided as supplementary material.

\subsection{Correlation}
\begin{figure}[ht]
    \centering
    \includegraphics[width=\linewidth]{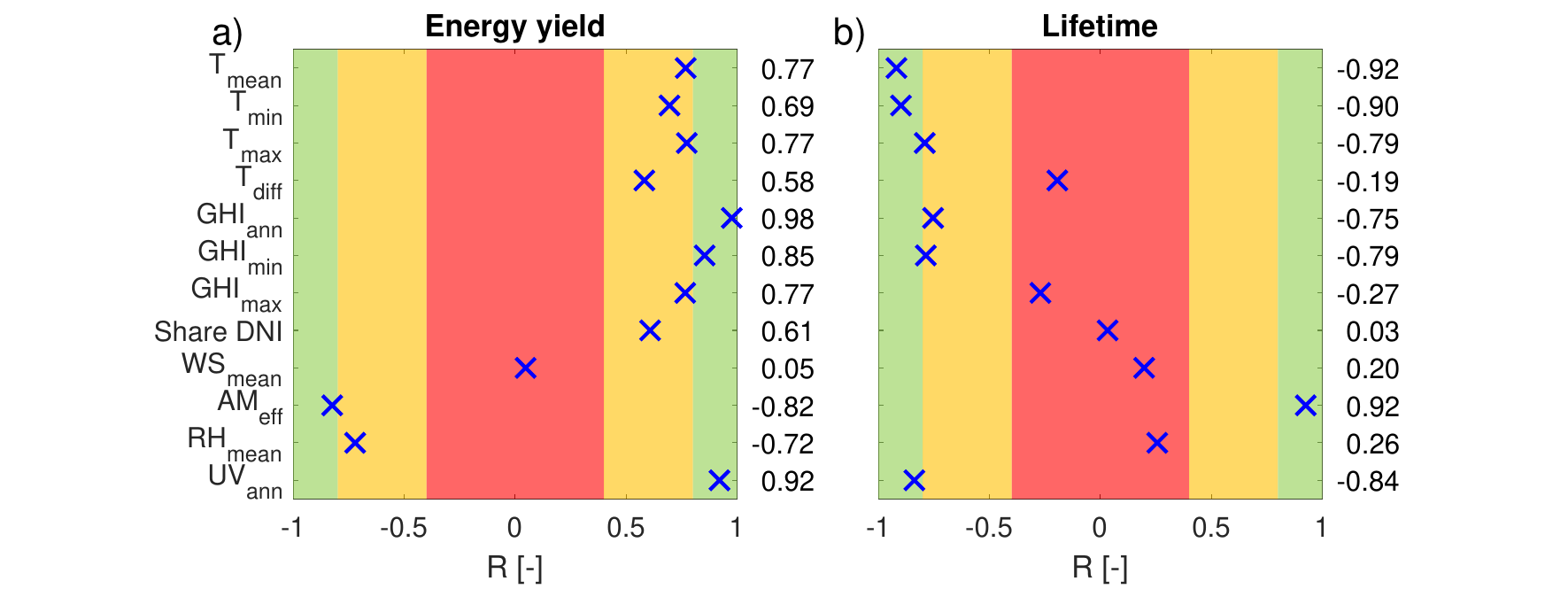}
    \caption{The correlation coefficients between all features and the energy yield (a) and module lifetime (b). Their values are on the right-hand side of both diagrams. The background colors indicate regions of no significant correlation (red), some significant correlation (orange), and strong significant correlation (green).}
    \label{fig:Correlation}
\end{figure}

To obtain an initial indication of feature relevance, Pearson correlation coefficients~\cite{benesty2009pearson} ($R$) are computed for all features. Figure~\ref{fig:Correlation} displays the correlations between each feature and the two targets: energy yield and module lifetime.
For energy yield, $\textrm{GHI}_{\textrm{ann}}$ exhibits the highest correlation, which is expected since it directly represents the total annual solar irradiation incident on the module.
For module lifetime, $\textrm{T}_{\textrm{mean}}$ shows the strongest correlation, reflecting the strong influence of temperature on degradation rates.  
Furthermore, $\textrm{AM}_{\textrm{eff}}$ demonstrates a correlation of similar magnitude but opposite sign compared to $\textrm{T}_{\textrm{mean}}$, which is consistent with the strong inverse relationship between these two features, as shown in Figure~\ref{fig:Targets}c.

Although the correlation coefficient offers direct insight into linear relationships between features and targets, relying solely on correlation for feature relevance has several well-known limitations, such as only accounting for linear trends and the influence of the range of observations~\cite{Janse2021ConductingPitfalls}. 
Therefore, more advanced ML-based techniques are employed to determine the feature importance, as described in the next section.

\section{Methodology}\label{sec:Methodology}
\begin{figure}[ht]
    \centering
    \includegraphics[width=\linewidth]{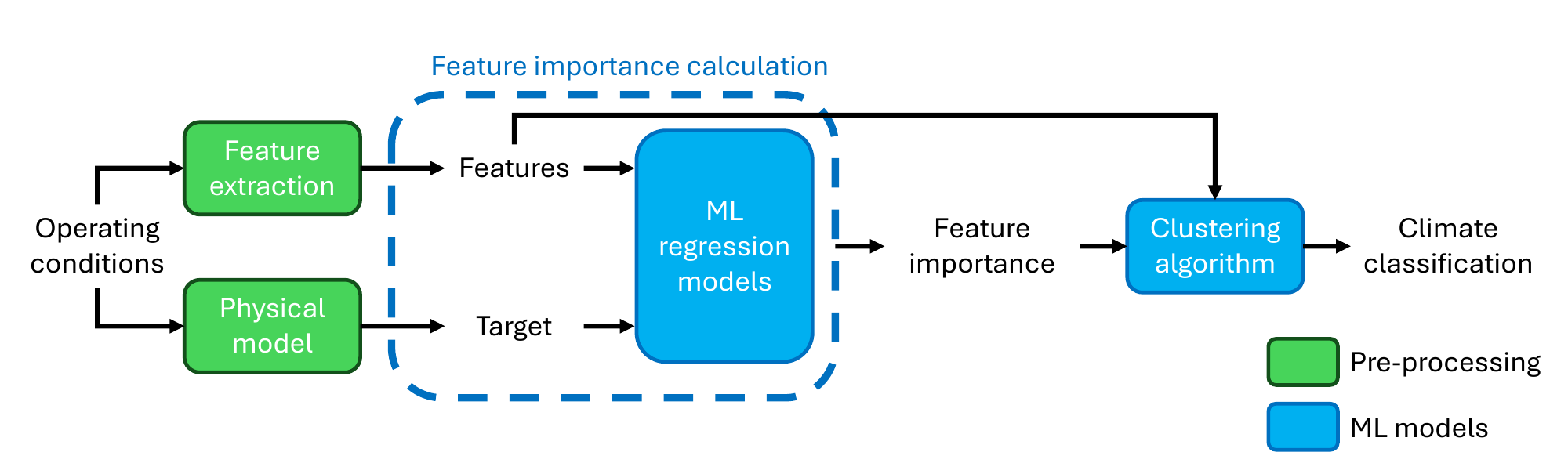}
    \caption{An overview of the methodology used for the climate classification. Features and targets are obtained from the operating conditions and the physical models. Feature importance is then determined using ML regression models and subsequently used as input for the clustering procedure.}
    \label{fig:OverviewMethodology}
\end{figure}

The climate classification is constructed through several methodological steps. An overview of the complete workflow is shown in Figure~\ref{fig:OverviewMethodology}. 
Following the pre-processing described in the previous section, a feature importance analysis is performed using multiple ML regression methods. 
These feature importance scores are then used as inputs to a clustering algorithm, which groups locations based on similarities in the most relevant environmental characteristics.

\subsection{Calculation of feature importance}
To quantify the importance of all features, a brute-force approach is employed, similar to the method proposed by Meinshausen~\textit{et al.}~\cite{Meinshausen2010StabilitySelection}. 
This approach incorporates five different ML regression methods, which are discussed later in this section. 
Using multiple ML methods allows for a more robust assessment of feature relevance, reducing dependence on any single model.

Instead of directly using all available features, the number of included features ($N_i$) is varied from 1 to the total number of features ($N_{\mathrm{feat}}$).
For each value of $N_i$, all possible feature combinations ($\binom{N_{\mathrm{feat}}}{N_i}$ in total) are evaluated, and the combination with the lowest error is selected. 
These selected sets represent the feature combinations that contain the most information for a given value of $N_i$.

For each feature $j$, an importance score $S_j$ is defined as

\begin{equation}\label{eq:ImportanceScore}
    S_j = \sum_{N_i}^{N_{feat}}\sum_{k}^{N_{models}}\frac{\epsilon_{max}}{N_i}\left(\frac{1}{\epsilon_k(N_i)}-\frac{1}{\epsilon_k(N_i-1)}\right)\cdot K(j,N_i,k),
\end{equation}
where $k$ denotes the ML method, $N_{models}$ is the total number of ML methods, $\epsilon_k(N_i)$ is the model error of method $k$ when using $N_i$ features, $\epsilon_{max}$ is the maximum error observed for a given target across all models. 
Including $\epsilon_{max}$ is important as it allows to fairly add the importance score of the different targets.
The function $K(j, N_i, k)$ is binary, equaling 1 if feature $j$ is selected by method $k$ at iteration $N_i$, and 0 otherwise.

Equation~\ref{eq:ImportanceScore} can be interpreted as assigning a performance-based boost to each selected feature at iteration $N_i$ for all models. The magnitude of this boost depends on the improvement in model performance relative to the previous iteration ($N_i - 1$) and is distributed evenly across all selected features. For completeness, $\epsilon_k(0)$ is defined to be infinite.

\subsection{Machine learning regression models}
Table~\ref{tab:MLregressionmodels} provides an overview of the ML regression models utilized in this work. 
The selected models span a range of regression types, including linear, kernel-based, spline-based, probabilistic, and neural network approaches. 
This diversity enables a robust and model-independent assessment of feature importance.

\begin{table}[ht]
    \centering
    \caption{An overview of the utilized ML regression methods that are used in this work.}
    \begin{tabular}{p{5cm}| p{2cm}| p{5cm}}
    \textbf{ML method} & \textbf{Reference} & \textbf{Modifications compared to default MATLAB settings} \\
    \hline
    Linear regression & \cite{Chatterjee1986InfluentalRegression} & \\
    Support Vector Machine & \cite{fan2005working, nash1994population} & \\
    Multivariate adaptive regression spline (MARS) &\cite{jekabsons2013adaptive, Friedman1991MultivariateSplines} & \\
    Neural network &\cite{Glorot2010UnderstandingNetworks, He2015DelvingClassification} & Changed activation function to sigmoid~\cite{Narayan1997TheLearning}\\
    Gaussian process regression &\cite{nash1994population} & \\
    \end{tabular}
    \label{tab:MLregressionmodels}
\end{table}

For each model, the dataset is split into 85\% training data and 15\% test data, and the Root Mean Square Error (RMSE) is used as the performance metric. Given the large number of feature combinations ($2^{12}$ in total), exhaustive hyperparameter optimization is computationally impractical. 
Instead, fixed hyperparameter settings are used, with minor adjustments as listed in Table~\ref{tab:MLregressionmodels}.

\subsection{Clustering technique}\label{sec:MethodologyClustering}
The final step in the climate classification is clustering, for which agglomerative hierarchical clustering is employed. 
Hierarchical clustering is chosen over alternative methods, such as k-means clustering~\cite{Hartigan1979AlgorithmAlgorithm}, because it enables the construction of nested sub-classifications within larger clusters in a systematic and transparent manner.
This hierarchical structure allows the level of classification detail to be easily adjusted, making the approach flexible for different applications.

As described by Murtagh~\textit{et al.}~\cite{Murtagh2012AlgorithmsOverview}, pairwise distances between all samples are first computed in the $N_{\mathrm{feat}}$-dimensional feature space using a predefined distance metric (here, Euclidean distance). 
These distances are then used to iteratively merge samples into clusters, resulting in a dendrogram that represents the hierarchical clustering structure.
A final classification is obtained by selecting an appropriate cut-off point in this dendrogram.

In hierarchical clustering, different linkage methods define how distances between clusters are computed. 
Common linkage methods include single, complete, average, centroid, median, weighted, and Ward linkage. In this work, the Ward linkage method~\cite{Ward1963HierarchicalFunction} is employed, as it produced the most balanced cluster structure among the tested linkage methods.
Ward linkage defines inter-cluster distances based on the increase in within-cluster variance upon merging clusters and can be expressed in terms of centroid distances under Euclidean assumptions. 
A comparison of alternative linkage methods is provided in~\ref{App:ComparisonDistanceMetrics}.
To ensure that differences in the most influential features contribute more strongly to the clustering, the feature values are weighted by their importance scores $S_j$ from Equation~\ref{eq:ImportanceScore} before constructing the dendrogram.

\section{Results}\label{sec:Results}
The methods and dataset described in the previous sections are now applied to construct the climate classification. 
We begin by examining the accuracy of the ML regression models, followed by the calculation of the feature importance scores. 
Finally, the clustering results and resulting climate classifications are presented.

\subsection{Accuracy of the regression models}
\begin{figure}[ht]
    \centering
    \includegraphics[width=\linewidth]{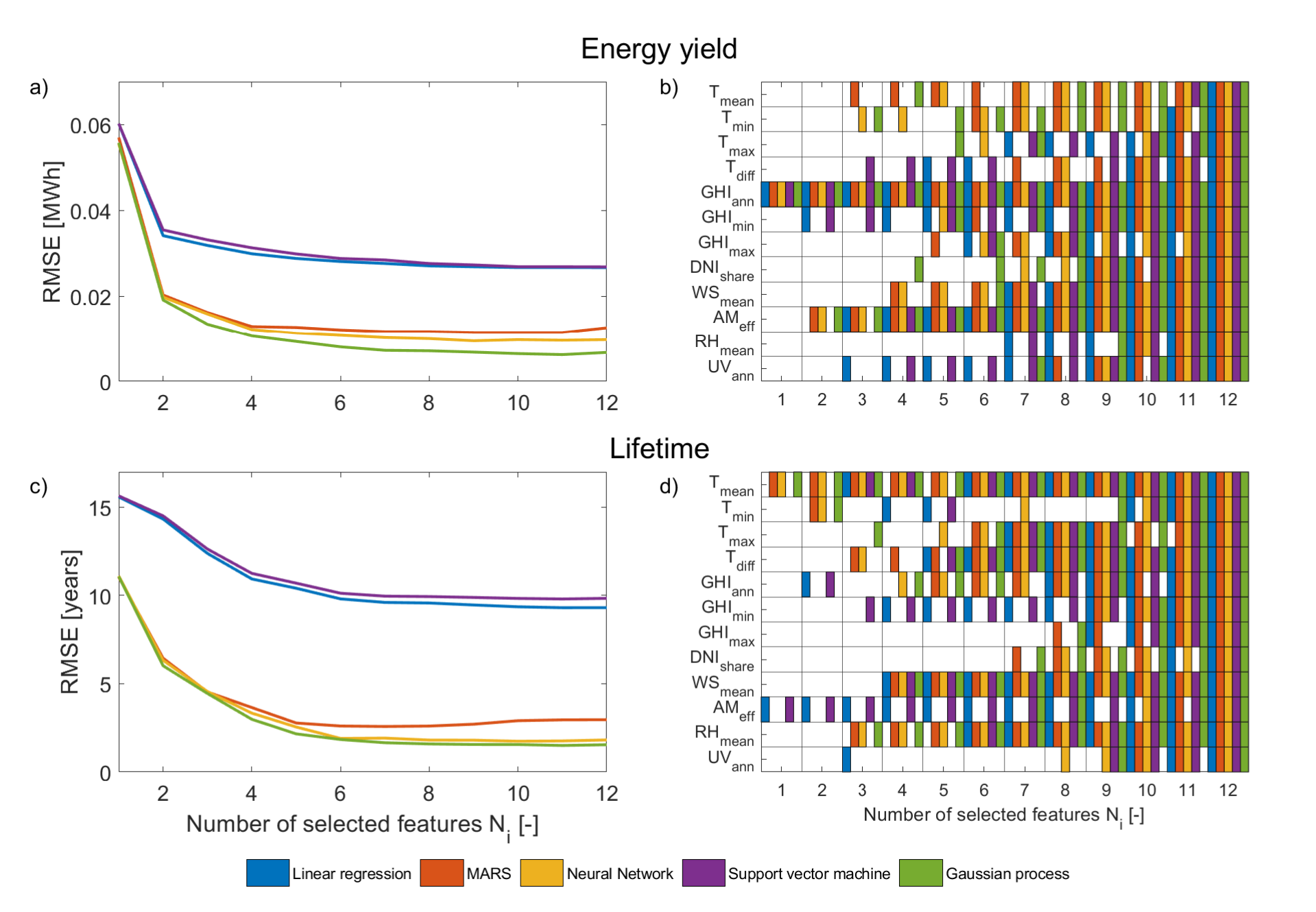}
    \caption{a) The highest accuracy of the different methods with number of selected features $N_i$ using the energy yield as target. b) The selected features by the different models for the energy yield as target. A colored box indicates that the method has selected a certain feature. c) The highest accuracy of the different methods with number of selected features $N_i$ using the lifetime as target. d) The selected features by the different models for the lifetime as target.}
    \label{fig:ResultsAccuracy}
\end{figure}

All methods listed in Table~\ref{tab:MLregressionmodels} are employed in the brute‑force feature‑selection approach. 
Figure~\ref{fig:ResultsAccuracy} shows the accuracy of each method for different number of features $N_i$, along with the selected features at each stage for both targets.
As expected, model performance improves as more features become available. 
For both targets, the RMSE decreases rapidly at low values of $N_i$ and begins to plateau around $N_i = 6$, indicating that six features capture most of the relevant information. 
For certain models, the RMSE increases again at $N_i = 12$, suggesting potential overfitting when all features are included.
Linear regression and Support Vector Machines yield the highest RMSE values, which is consistent with their limited ability to capture non‑linear relationships. 
The remaining models, which can model non‑linear trends, achieve considerably higher accuracy, with Gaussian Process Regression showing the best overall performance, reaching RMSE values of 0.007~\unit{\mega\watt\hour} and 1.5 year for the energy yield and lifetime target, respectively.

Figure~\ref{fig:ResultsAccuracy}b) and d) show that some features are utilized more often than others.
For the energy yield target, the $\textrm{GHI}_{\textrm{ann}}$ and $\textrm{AM}_{\textrm{eff}}$ are most often selected, suggesting that these features contain the highest information.
Similarly, $\textrm{T}_{\textrm{mean}}$ and $\textrm{RH}_{\textrm{mean}}$ are most frequent selected when the lifetime is used as target.
These results will be used to quantify the importance of each feature.

\subsection{Feature importance}
\begin{figure}[ht]
    \centering
    \includegraphics[width=\linewidth]{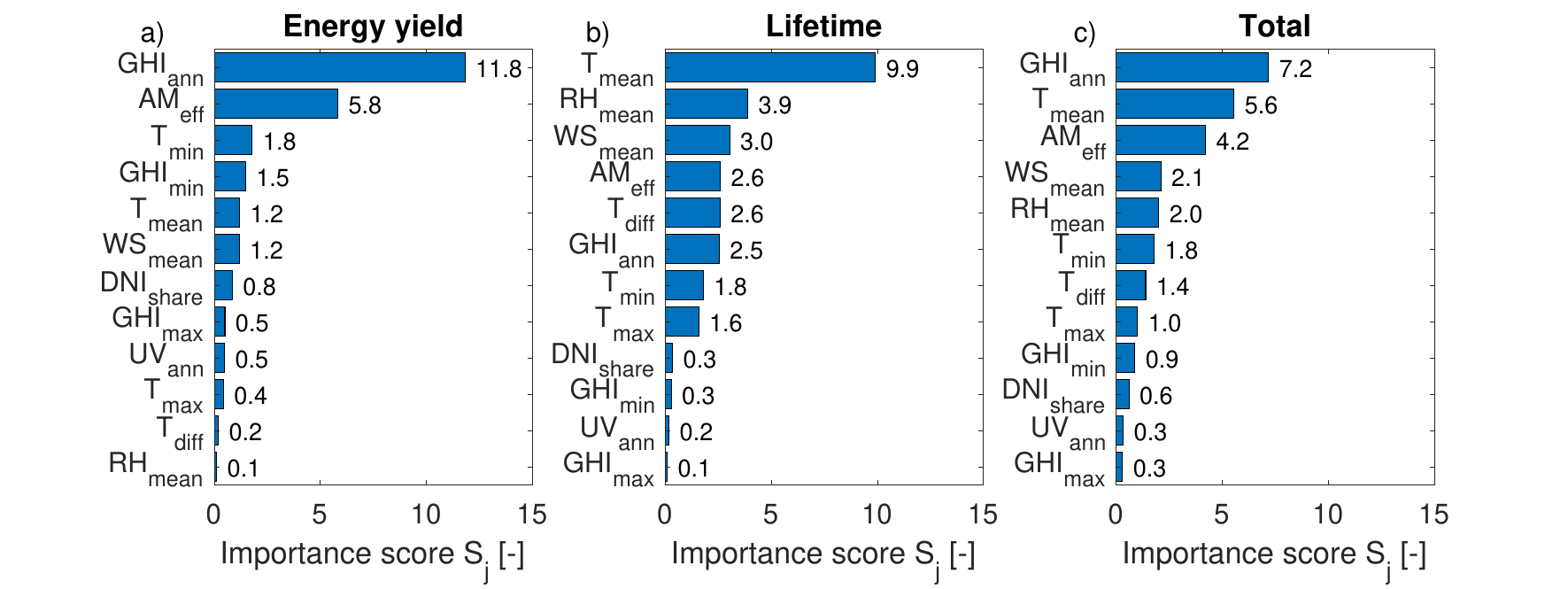}
    \caption{The importance score of all features considering the energy yield (a) or lifetime (b) as target. The total feature importance score (c) is obtained by taking the average of the two.}
    \label{fig:FeatureScore}
\end{figure}

The results from Figure~\ref{fig:ResultsAccuracy} together with Equation~\ref{eq:ImportanceScore} are used to compute the importance scores $S_j$ for all features for both targets. 
Figures~\ref{fig:FeatureScore}a and~\ref{fig:FeatureScore}b show the resulting feature importance scores for the energy‑yield and lifetime targets, respectively. 
For both targets, the features with the highest scores correspond well with the most frequently selected features shown previously in Figure~\ref{fig:ResultsAccuracy}.

To obtain a unified importance measure that reflects both energy‑yield and lifetime considerations, the average of the two target‑specific scores is taken, as shown in Figure~\ref{fig:FeatureScore}c. 
The features $\textrm{GHI}_{\textrm{ann}}$ and $\textrm{T}_{\textrm{mean}}$ emerge as the two most influential parameters, each topping the importance ranking for the energy yield and lifetime targets, respectively. 
These combined scores (Figure~\ref{fig:FeatureScore}c) are used as input for the clustering procedure.

\subsection{Clustering \& Classification}
The method described in Section~\ref{sec:MethodologyClustering} is applied to construct the climate classification using the feature importance scores from Figure~\ref{fig:FeatureScore}c, thereby incorporating both energy‑yield and lifetime aspects. 
It should be noted that using the feature importance scores from either Figure~\ref{fig:FeatureScore}a or~\ref{fig:FeatureScore}b would allow for classifications focused solely on energy yield or lifetime, respectively. 
Corresponding classifications and analyses for these individual targets are provided in \ref{App:TargetClassifications}.

\begin{figure}[ht]
    \centering
    \includegraphics[width=\linewidth]{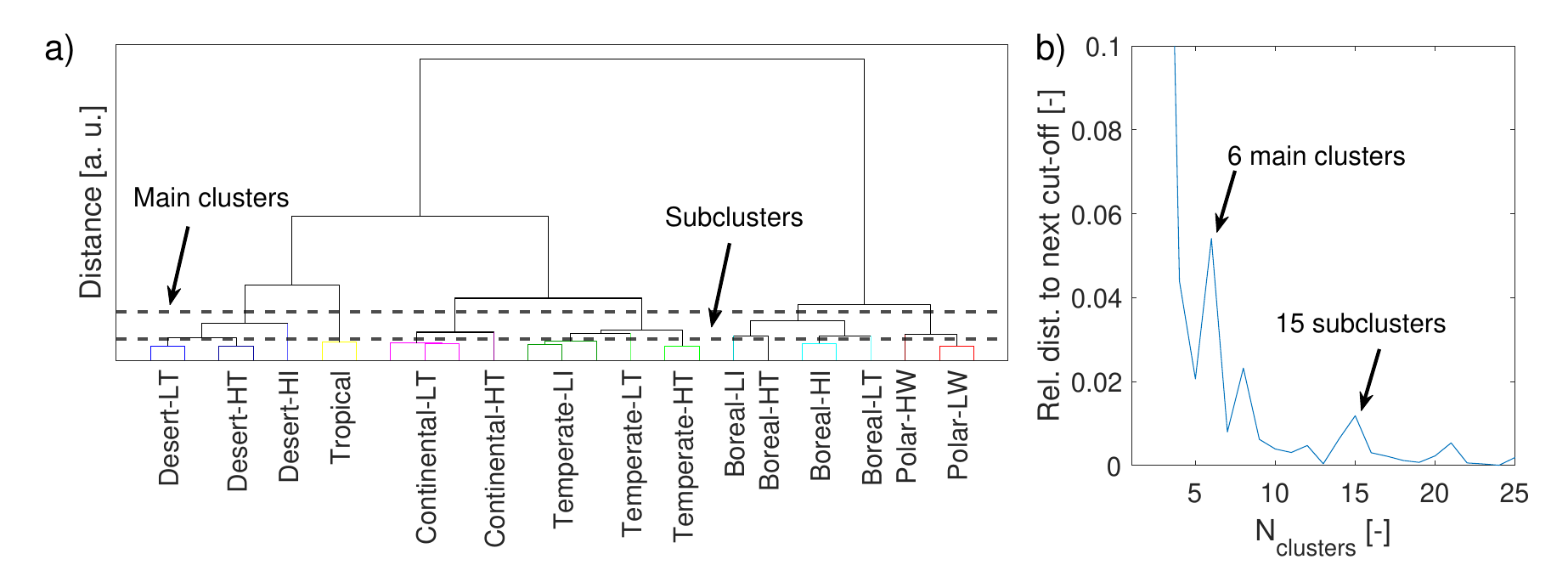}
    \caption{a) The dendrogram that is created based on the distance map and the corresponding cluster labels. b) The relative distance to the next cut-off is plotted for each value of $N_{clusters}$. The values of $N_{clusters}=2$ and $N_{clusters} = 3$ (0.51 and 0.23, respectively) are not visible, since the image is cropped to clearly show the meaningful part of the graph. This figure can be used to determine where to cut-off the dendrogram.}
    \label{fig:Dendrogram}
\end{figure}

Figure~\ref{fig:Dendrogram}a shows the dendrogram constructed from the weighted distance map. 
While the figure displays only the first 25 clusters for readability, the hierarchical structure continues until each data point forms its own cluster. 
To derive a practical climate classification, a cut-off threshold must be chosen to determine the number of clusters ($N_{\mathrm{cluster}}$). 
However, selecting the optimal number of clusters is not straightforward, as no universally accepted metric exists for this purpose~\cite{Mirkin2011ChoosingClusters}.

In this work, we use the distances in the dendrogram to determine the optimal value of $N_{clusters}$. 
Figure~\ref{fig:Dendrogram}b shows the relative distance between successive cut-off points for different values of $N_{clusters}$. 
Several local maxima can be observed; these represent points in the dendrogram where the distance between adjacent cut-offs is relatively large, indicating potentially meaningful locations at which to divide the hierarchy.
Although the edge case of $N_{\mathrm{clusters}} = 2$ exhibits the largest relative distance (0.51), this option is disregarded, as a classification with only two clusters would not provide useful differentiation.

The first local maximum occurs at $N_{\mathrm{clusters}} = 6$, which is therefore selected for defining the 6 main climate clusters ($N_{\mathrm{main}}$). 
To obtain finer granularity, subclusters ($N_{\mathrm{sub}}$) are also introduced. 
We choose $N_{\mathrm{sub}} = 15$, corresponding to the first local maximum for which $N_{\mathrm{clusters}} > 2 \cdot N_{\mathrm{main}}$. 
The resulting two cut-off levels used for classification are indicated by dashed lines in Figure~\ref{fig:Dendrogram}.
Also, the corresponding cluster labels are plotted.

One thing that should be realized is that the created dendrogram is sensitive to small changes in the computed distances.
This means that the number of optimal clusters can change for small changes in the feature importance score, as illustrated in~\ref{App:SensitivityAnalysis}.
Future research effor can be spent on analyzing the impact of small changes to the classification.

\begin{sidewaysfigure}[p]
    \centering
    \includegraphics[width=\linewidth]{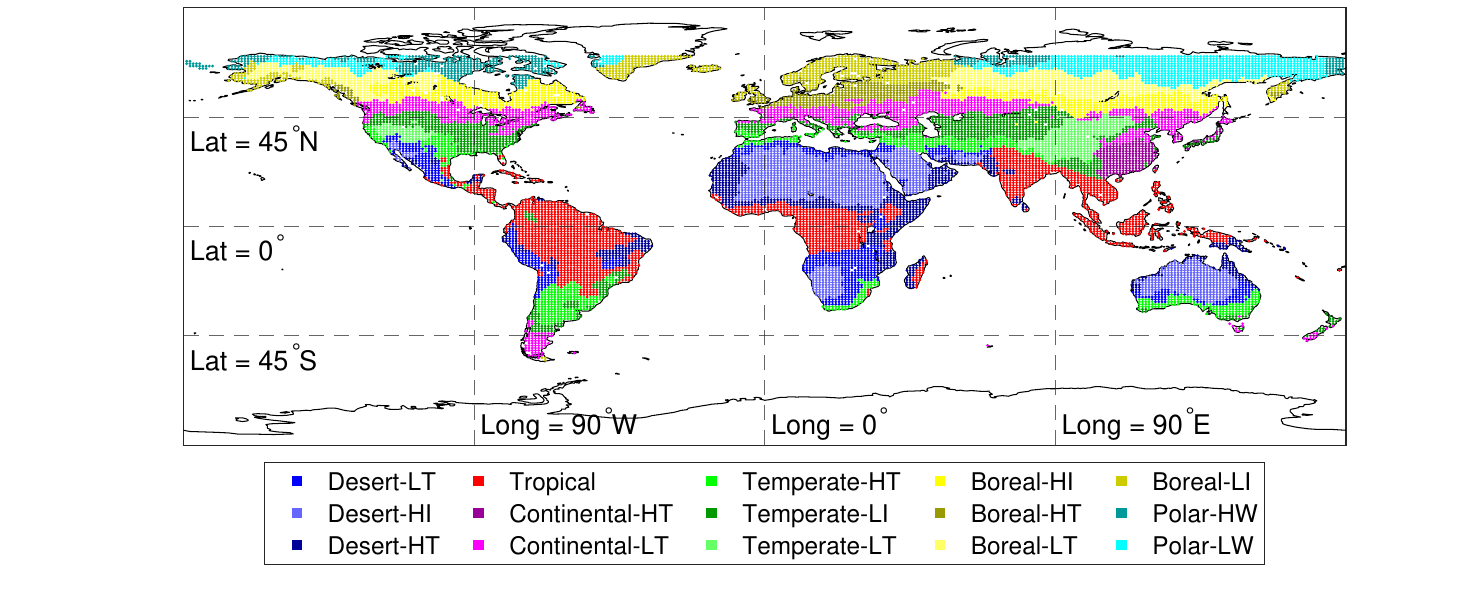}
    \caption{The final climate classification using $N_{main}=6$ and $N_{sub}=15$. The main climates are labelled as Tropical, Desert, Continental, Temperate, Boreal, and Polar. Subcluster are further distinguished by irradiance (I), temperature (T), or wind speed (W) levels, denoted as low (L) and high (H).}
    \label{fig:ClimateClassification}
\end{sidewaysfigure}

Figure~\ref{fig:ClimateClassification} presents the final climate classification using $N_{\mathrm{main}} = 6$ and $N_{\mathrm{sub}} = 15$. 
The six main climates are labeled Tropical, Desert, Continental, Temperate, Boreal, and Polar, closely aligning with the terminology used in the KGPV classification~\cite{Ascencio-Vasquez2019MethodologyPerformance}. 
The subclusters within each main climate are further distinguished by levels of irradiance (I), temperature (T), or wind speed (W), denoted as low (L) or high (H).

As an example, the Desert climate is subdivided into three zones: low temperature (LT), high temperature (HT), and high irradiance (HI). 
It should be noted that these subcluster labels express the relative feature values within each main climate.
A detailed explanation of the subcluster labeling procedure is provided in~\ref{App:LabelignSubclusters}. 
Furthermore, an overview comparing this climate classification with previously developed classifications is presented in~\ref{App:OverlapExisting}.

\begin{figure}[ht]
    \centering
    \includegraphics[width=\linewidth]{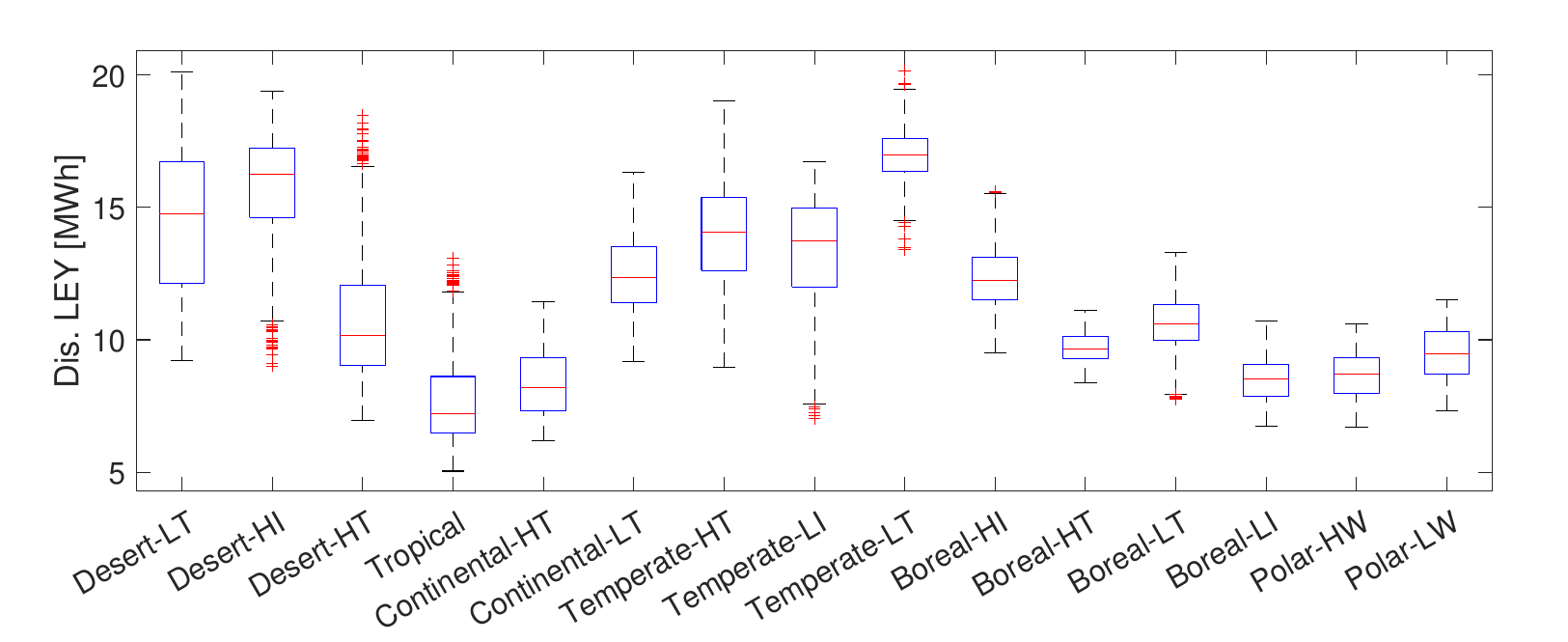}
    \caption{The box plot distribution of the discounted lifetime energy yield for the different climate zones.}
    \label{fig:BoxPlotsLEY}
\end{figure}

To compare the performance of PV systems across the different clusters, Figure~\ref{fig:BoxPlotsLEY} presents box plots of the discounted Lifetime Energy Yield (LEY) for each climate zone. 
The discounted LEY assumes a discount rate of 7\%~\cite{IEA2020_ProjectedElectricity}, and its mathematical definition is given in~\ref{App:DefinitionDisLEY}.

The results show that the highest LEY values occur in the temperate‑LT climate, where high energy yields coincide with long module lifetimes. 
Additionally, across all main climates, the LT subclusters consistently outperform their HT counterparts. 
This trend is expected, as lower temperatures reduce degradation rates and thereby extend module lifetime.

The climate classification in Figure~\ref{fig:ClimateClassification}, together with the LEY distributions in Figure~\ref{fig:BoxPlotsLEY}, enables several practical applications.  
First, the classification can guide optimized cell‑ or BoM‑design strategies, allowing PV manufacturers to tailor module designs to specific climate zones.  
Second, site‑allocation decisions for PV systems can benefit from the classification, as it highlights the regions offering the highest lifetime performance.  
Lastly, the classification facilitates fair comparisons between different PV technologies.
By selecting a representative set of locations, one for each climate zone, global performance comparisons can be conducted without the need to evaluate a large number of sites.
To facilitate this comparison, \ref{App:RepresentativeLocations} provides an overview of representative locations for each climate zone.

\section{Conclusion}\label{sec:Conclusion}
This work presents a climate classification that utilizes multiple ML methods and incorporates both energy‑yield and lifetime aspects. 
First, a dataset is constructed containing twelve environmental features and two performance targets. 
To correct for the non‑uniform geographical distribution of the raw data, spatial interpolation is applied to obtain a uniformly distributed global dataset. 
A brute‑force approach using five different ML regression methods is then employed to quantify the importance of each feature, and the resulting feature importance scores are subsequently used as inputs for hierarchical clustering.

The analysis shows that linear regression and Support Vector Machines yield the lowest performance, as both methods are limited to modeling predominantly linear relationships.
The gaussian process reaches the highest accuracy, obtaining RMSE values of 0.007~\unit{\mega\watt\hour} and 1.5 years for the energy yield and lifetime, respectively.
Furthermore, the performance of all ML methods saturates once more than six features are included, indicating that the majority of relevant information is captured within the top six features. 
The most influential features for the energy‑yield and lifetime targets are $\textrm{GHI}_{\textrm{ann}}$ and $\textrm{T}_{\textrm{mean}}$, respectively.

Based on the dendrogram structure, six main climate clusters and fifteen subclusters are defined, corresponding to the largest relative gaps in the hierarchy. 
The main climates are labeled Tropical, Desert, Continental, Temperate, Boreal, and Polar, and are further differentiated by relatively high or low values of temperature, irradiance, or wind speed. 
Finally, the discounted lifetime energy yield (LEY) is evaluated across all clusters, with the highest values observed in the Temperate‑LT climate zone.
These results enable various applications, including optimized PV module and BoM design, improved PV system siting decisions, and more efficient comparative studies across PV technologies.

\section*{Declaration of Generative AI and AI-assisted technologies in the writing process}
During the preparation of this work the author used CoPilot in order to paraphrase sentences and improve the language and readability. 
After using this tool/service, the author reviewed and edited the content as needed and takes full responsibility for the content of the publication.

\section*{Supporting Information}
The dataset and the software of the methods can be accessed on a 4TU dataset via this link: \url{https://doi.org/10.4121/c7e04a2d-123a-40a9-a4e7-f260f8f23e5c}

\appendix
\section{Definition of features}
\label{App:FeatureDefinitions}
\setcounter{figure}{0} 
Table~\ref{tab:Features} provides an overview of all features considered in this work. 
These features are derived from the hourly weather data obtained from Meteonorm~\cite{remund2020meteonorm}, which includes ambient temperature, global horizontal irradiance (GHI), direct normal irradiance (DNI), diffuse horizontal irradiance (DHI), sun position, wind speed (WS), and relative humidity (RH).

Most features ($\textrm{T}_{\textrm{mean}}$, $\textrm{T}_{\textrm{min}}$, $\textrm{T}_{\textrm{max}}$, $\textrm{GHI}_{\textrm{ann}}$, $\textrm{GHI}_{\textrm{min}}$, $\textrm{GHI}_{\textrm{max}}$, $\textrm{WS}_{\textrm{mean}}$, and $\textrm{RH}_{\textrm{mean}}$) are obtained simply by taking the annual mean, monthly minimum, monthly maximum, or sum of the corresponding hourly values. However, several features require more complicated definitions.

The feature $\textrm{T}_{\textrm{diff}}$ represents the average daily temperature fluctuation and is defined as

\begin{equation}
    \textrm{T}_{\textrm{diff}}=\frac{\sum_i^{N_{days}}\left(\max(T_{amb}(t\in hours_i))-\min(T_{amb}(t\in hours_i))\right)}{N_{days}},
\end{equation}
where $N_{days}$ are the number of days in the year, $T_{amb}(t)$ is the ambient temperature at hour $t$, $hours_i$ represent all hours within day $i$.

$\textrm{DNI}_{\textrm{share}}$ is defined as the share of direct irradiance contributing to the total irradiance, which is defined as

\begin{equation}
    \textrm{DNI}_{\textrm{share}} = \frac{\sum_t\left(DNI(t)\cdot \cos(\phi_{sun}(t))\right)}{\sum_t GHI(t)},
\end{equation}
where $DNI(t)$, $GHI(t)$ and $\phi_{sun}(t)$ are direct normal irradiance, global horizontal irradiance, and sun altitude, respectively, at time $t$.

$\textrm{AM}_{\textrm{eff}}$ is the effective air mass at which the irradiance is received, which is defined as

\begin{equation}
    \textrm{AM}_{\textrm{eff}} = \frac{\sum_t\left(\frac{1}{\cos(\phi_{sun}(t))}\cdot GHI(t)\right)}{\sum_t GHI(t)}
\end{equation}

Finally, $\textrm{UV}_{\textrm{ann}}$ denotes the annual received ultraviolet irradiance. 
This quantity is computed directly by the PVMD Toolbox and corresponds to all incident irradiance with wavelengths below 400~\unit{\nano\meter}.

\section{Validation lifetime calculation}
\label{App:LifetimeValiation}
\setcounter{figure}{0}
Figure~\ref{fig:Targets} shows the geographical distribution of the simulated energy yield and module lifetime. 
In some locations, the simulated lifetime exceeds the commonly expected module lifetime of 30/40 years. 
These high values arise from the definition used in this study, where lifetime corresponds to the time required for the module performance to decrease to 80\% of its initial value~\cite{Blom2026CombiningModules}. 
Because temperature is the dominant stress factor influencing degradation, regions with consistently low temperatures, such as Nordic countries, naturally exhibit lower degradation rates and therefore longer lifetimes. 
Psimopoulos~\textit{et al.}~\cite{Psimopoulos2024PerformanceScandinavia} report that PV modules in Sweden retain more than 90\% of their initial performance after 30 years. 
Linearly extrapolating these results suggests that reaching the 80\% threshold could indeed require around 60 years.

\begin{figure}[ht]
    \centering
    \includegraphics[width=0.6\linewidth]{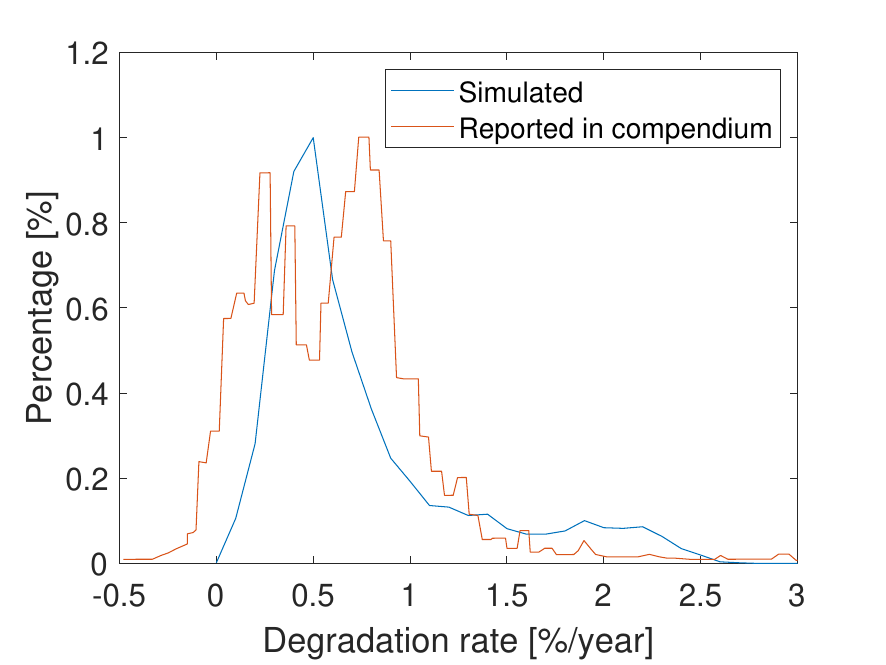}
    \caption{The distribution of degradation rates obtained from the degradation model and collected by the compendium of degradation rates~\cite{Jordan2016CompendiumRates}.}
    \label{fig:validationLifetime}
\end{figure}

To obtain a broader validation, reported degradation rates from the global compendium compiled by Jordan~\textit{et al.}~\cite{Jordan2016CompendiumRates} are used for comparison. 
Figure~\ref{fig:validationLifetime} shows the relative distribution of degradation rates predicted by the degradation model alongside those collected in the compendium.

Although there are noticeable differences between the simulated and reported values, the overall range of degradation rates is comparable. 
This indicates that the simulated lifetimes are not unrealistic. The remaining discrepancies can be attributed to differences in module technologies, installation characteristics, and geographical sampling between the model and the reported datasets.

\section{Considered module}
\label{App:ConsideredModule}
\setcounter{figure}{0}
The energy yield and module lifetime targets from the dataset are simulated with a module also used in previous work~\cite{Blom2026CombiningModules}.
It is a monofacial silicon heterojunction module, consisting of 144 half-cut cells with a G12 wafer size, and is equipped with an EVA encapsulant and PET backsheet.
Figure~\ref{fig:ModuleDesign} shows the optical and electrical module performance.
The full details (and the considered parameters of the lifetime simulations) can be found in the previous work~\cite{Blom2026CombiningModules}.

\begin{figure}[ht]
    \centering
    \includegraphics[width=0.8\linewidth]{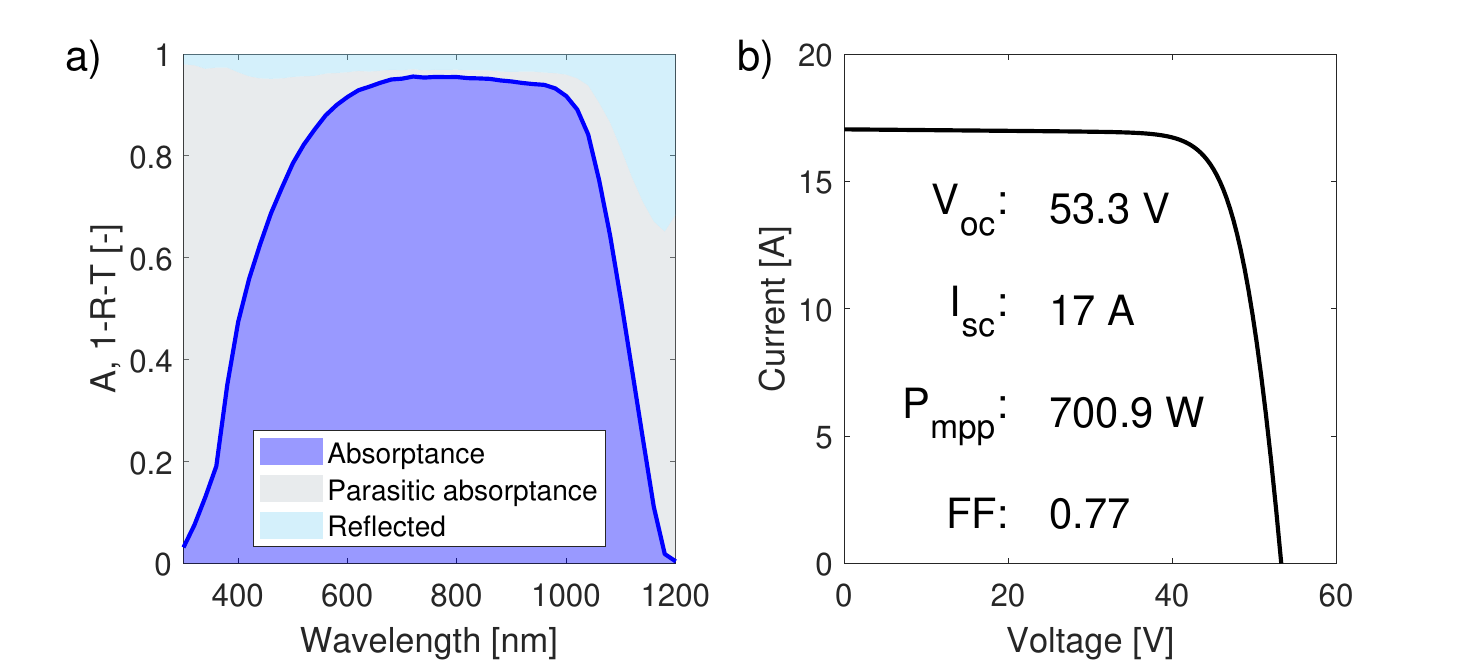}
    \caption{The optical performance (a) and electrical performance (b) of the considered module at standard test conditions.}
    \label{fig:ModuleDesign}
\end{figure}

For each location, the module tilt and azimuth are chosen that maximize the received irradiance.
These values can be found in the supplementary information.




\section{Comparison distance metrics}
\label{App:ComparisonDistanceMetrics}
\setcounter{figure}{0}
To construct the dendrogram in Figure~\ref{fig:Dendrogram}a), both a distance metric between samples and a linkage method defining distances between clusters must be specified.
In this work, Euclidean distance is used as the underlying distance metric, while different linkage methods are evaluated.

The MATLAB environment~\cite{MatlabHelpCenterLinkage} provides several linkage methods, which are summarized in Table~\ref{tab:DistanceMetrics}. 
For all definitions, $d(r,s)$ denotes the distance between clusters $r$ and $s$, $N_r$ and $N_s$ are the number of samples in clusters $r$ and $s$, respectively, and $||x - y||_2$ denotes the Euclidean distance between two samples. It should be noted that the median and weighted linkage methods are defined recursively, where cluster $r$ is formed by merging clusters $p$ and $q$.

\begin{table}[ht]
    \centering
    \caption{Linkage methods used in hierarchical clustering and their corresponding definitions. For all equations, $d(r,s)$ is the distance between clusters $r$ and $s$, $N_r$ and $N_s$ are the number of points in cluster $r$ and $s$, and $||x-y||_2$ is the Euclidean distance between two points. It should be noted that the median and weighted metric are recursive metrics, where cluster $r$ is created by combining cluster $p$ and cluster $q$.}
    \begin{tabular}{p{2cm}| p{11cm}}
    \textbf{Metric} & \textbf{Equation}  \\
    \hline
    average & $d(r,s) = \frac{1}{N_r\cdot N_s}\sum_{i=1}^{N_r}\sum_{j=1}^{N_s}||x_{ri}-x_{sj}||_2$\\
    centroid & $d(r,s) = ||\bar{x}_r-\bar{x}_s||_2$\\
    complete & $d(r,s) = \max(||x_{ri}-x_{sj}||_2),i\in (1,...,N_r), j\in (1,...,N_s)$\\
    median & $d(r,s) = ||\tilde{x}_r-\tilde{x}_s||_2, \tilde{x_r} = \frac{1}{2}(\tilde{x}_p+\tilde{x}_q)$\\
    single & $d(r,s) = \min(||x_{ri}-x_{sj}||_2),i\in (1,...,N_r), j\in (1,...,N_s)$\\
    ward & $d(r,s) = \sqrt{\frac{2\cdot N_r\cdot N_S}{N_r+N_s}}||\bar{x}_r-\bar{x}_s||_2$\\
    weighted & $d(r,s) = \frac{d(p,s)+d(q,s)}{2}$\\
    \end{tabular}
    \label{tab:DistanceMetrics}
\end{table}

\begin{figure}[ht]
    \centering
    \includegraphics[width=\linewidth]{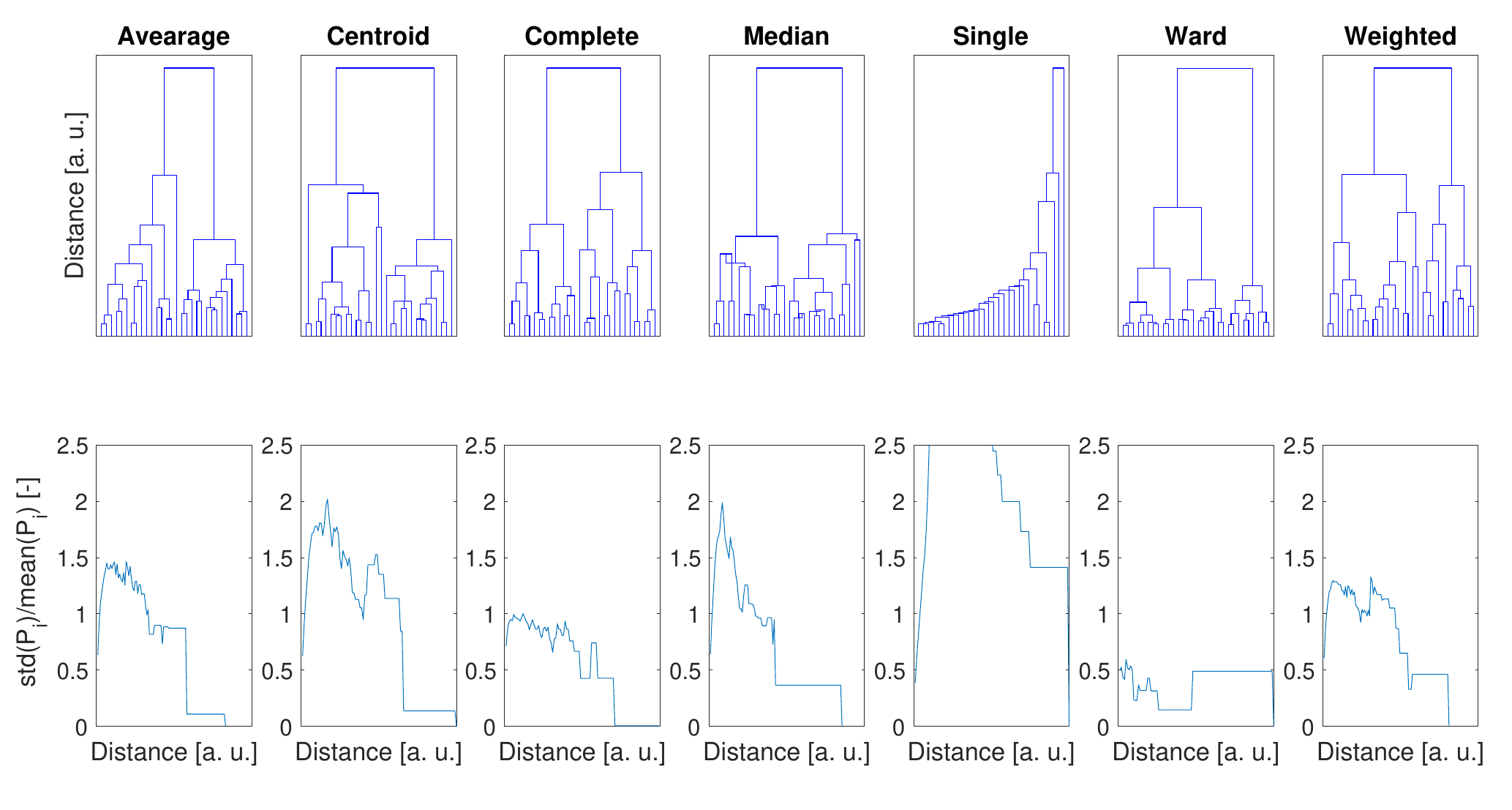}
    \caption{Comparison of hierarchical clustering results using different linkage methods. The top figures show the resulting dendrograms. The figures below show the standard deviation of points in the cluster over the average points in the cluster, indicating the spread of all clusters.}
    \label{fig:distanceMetrics}
\end{figure}

The choice of linkage method strongly influences the structure of the dendrogram and therefore affects the resulting climate classification.
The top plots of Figure~\ref{fig:distanceMetrics} illustrate how different distance metrics alter the dendrogram, demonstrating that its shape is highly sensitive to the metric used.
In this work, our objective is to identify a linkage method that produces a dendrogram in which data points are distributed as evenly as possible across clusters. 
Such a metric is desirable because it avoids classifications in which some climate zones correspond to only very small regions.

To quantify the uniformity of the distribution, we compute the standard deviation of the number of points per cluster normalized by the mean number of points per cluster across all levels of the dendrogram, written as $\frac{\mathrm{std}(P_i)}{\mathrm{mean}(P_i)}$, where $P_i$ denotes the number of points in cluster $i$. A lower value of this ratio indicates a more even distribution of data points across clusters.
The bottom panels of Figure~\ref{fig:distanceMetrics} show the values of $\frac{\mathrm{std}(P_i)}{\mathrm{mean}(P_i)}$ for all considered linkage methods. 
As observed, the Ward linkage consistently yields the lowest ratio across the entire dendrogram. 
Therefore, Ward linkage is selected for this work, as it produces the most uniform distribution of data points among clusters.

\section{Classifications for separate targets}
\label{App:TargetClassifications}
\setcounter{figure}{0}
The clustering and classification procedure can also be applied when considering only the energy‑yield or lifetime target. 
In this case, the feature values are weighted using the importance scores shown in Figures~\ref{fig:FeatureScore}a and~\ref{fig:FeatureScore}b, respectively. 
This weighting alters the resulting dendrogram and therefore the climate classification.
Figures~\ref{fig:differentTargetsDendrogram}a and~\ref{fig:differentTargetsDendrogram}c present the dendrograms obtained when considering only the energy yield or lifetime feature importance scores, respectively. 
The corresponding relative distances between successive cut‑off points in the dendrogram, used to determine the optimal number of clusters, are shown in Figures~\ref{fig:differentTargetsDendrogram}b and~\ref{fig:differentTargetsDendrogram}d.

\begin{figure}[ht]
    \centering
    \includegraphics[width=\linewidth]{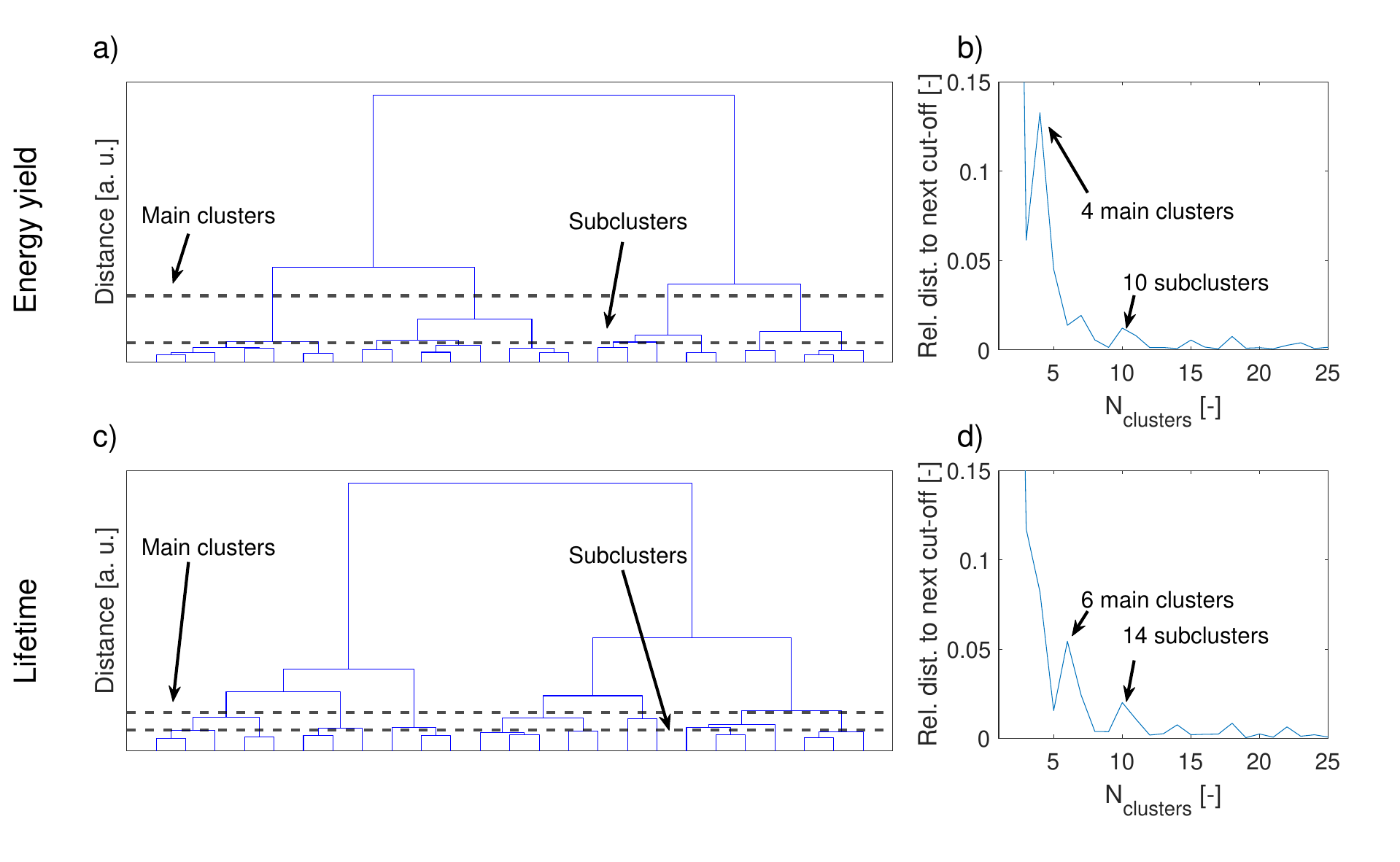}
    \caption{The different dendrogram and cluster selection when only considering the energy yield or lifetime. (a) and (b) show the dendrogram and relative distances for the energy yield target, finding local maxima for 4 and 10 clusters.(c) and (d) show the dendrogram and relative distances for the lifetime target, finding local maxima for 6 and 14 clusters.}
    \label{fig:differentTargetsDendrogram}
\end{figure}

\begin{figure}[h!]
    \centering
    \includegraphics[width=\linewidth]{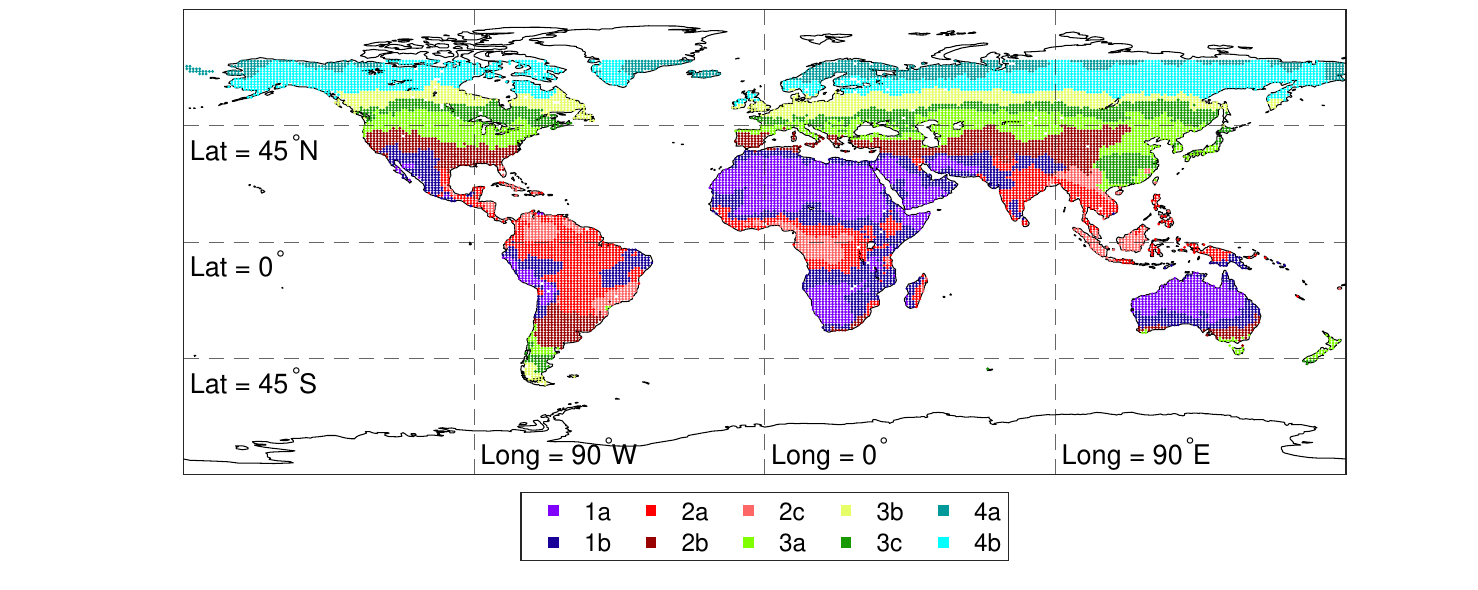}
    \caption{The climate classification when considering the feature scores for the energy yield targets. Main clusters are indicated with 1, 2, 3, and 4. Subclusters are indicated with a, b, and c to avoid confusion with the presented classification.}
    \label{fig:differentTargetsClassificationEnergyYield}
\end{figure}

\begin{figure}[h!]
    \centering
    \includegraphics[width=\linewidth]{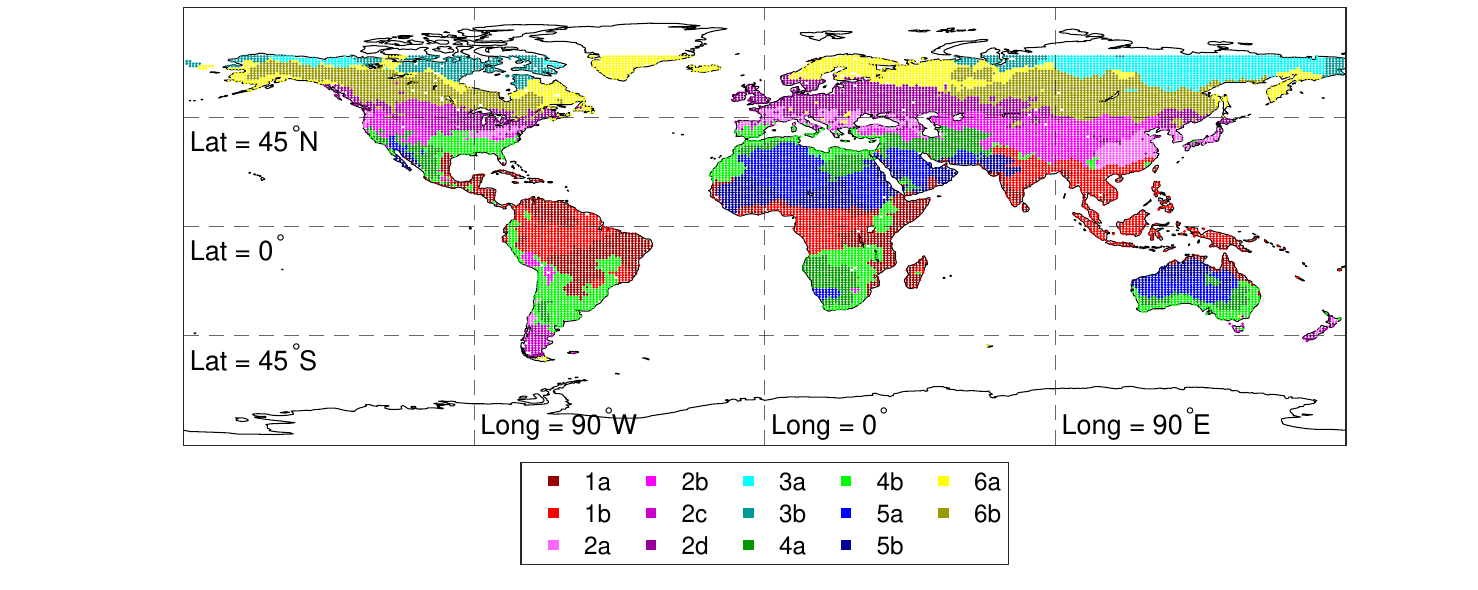}
    \caption{The climate classification when considering the feature scores for the lifetime targets. Main clusters are indicated with 1, 2, 3, 4, 5, and 6. Subclusters are indicated with a, b, c, and d to avoid confusion with the presented classification.}
    \label{fig:differentTargetsClassificationLifetime}
\end{figure}

Applying the same selection criteria as in the main analysis yields 4 main clusters and 10 subclusters for the energy‑yield‑only classification, and 6 main clusters and 14 subclusters for the lifetime‑only classification. 
The resulting classifications are shown in Figures~\ref{fig:differentTargetsClassificationEnergyYield} and~\ref{fig:differentTargetsClassificationLifetime}, respectively.

For clarity, the main clusters in these two alternative classifications are labeled numerically (1, 2, 3, \ldots), and subclusters alphabetically (a, b, c, \ldots). 
Using similar climate‑zone names as in the main study could lead to confusion, as the resulting regions differ substantially. 
When only an energy‑yield‑based or lifetime‑based classification is required for specific applications, future work may focus on assigning meaningful descriptive labels to these clusters.

\section{Sensitivity to changes in feature importance score}
\label{App:SensitivityAnalysis}
\setcounter{figure}{0}
The dendrogram obtained from hierarchical clustering is sensitive to small perturbations in the computed distances.
To illustrate this sensitivity, we reproduce Figure~\ref{fig:Dendrogram}b using slightly modified feature importance scores.
Specifically, a random value drawn uniformly from the interval $[-0.5,\,0.5]$ is added to each feature importance score.

\begin{figure}[ht]
    \centering
    \includegraphics[width=\linewidth]{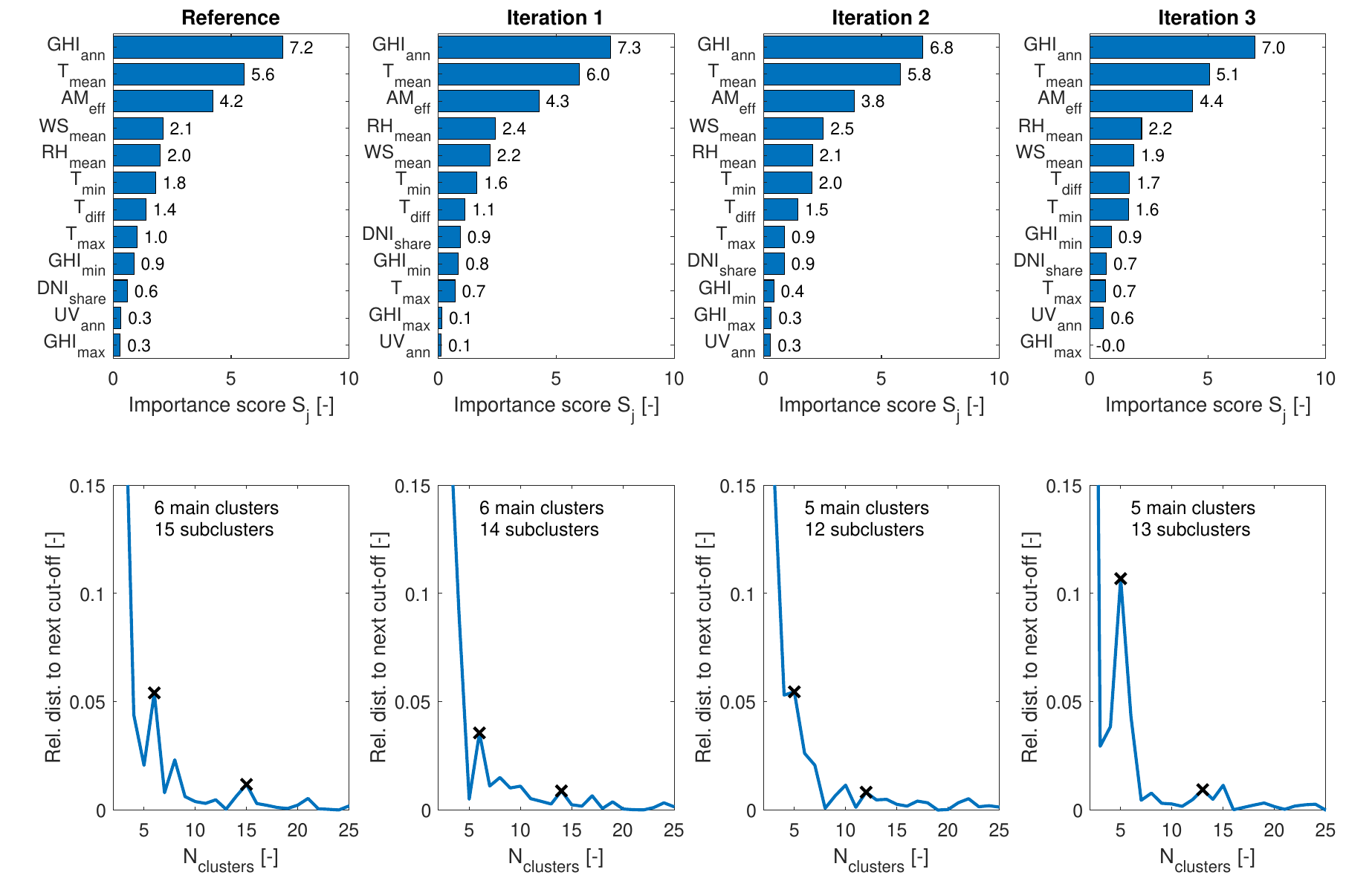}
    \caption{Sensitivity analysis for determining the optimal number of clusters. The top panels show the feature importance scores for the reference case (left) and three perturbed cases. 
    The bottom panels show the corresponding dendrogram distances and the resulting optimal numbers of main clusters and subclusters.}
    \label{fig:SensitivityAnalysis}
\end{figure}

Figure~\ref{fig:SensitivityAnalysis} shows the resulting optimal numbers of clusters and subclusters for three independent realizations of this perturbation.
Although these small changes do not alter the relative ranking of the three most important features, they can still affect the structure of the dendrogram and, consequently, the identified optimal number of clusters.
This highlights the sensitivity of the clustering outcome to minor variations in the feature importance scores.

\section{Labeling of subclusters}
\label{App:LabelignSubclusters}
\setcounter{figure}{0}
As explained in the main text, the subclusters are denoted with relative high (H) or low (L) values in temperature (T), irradiance (I), or wind speed (W).
To justify the labeling of the subclusters, Figure~\ref{fig:explainLabeling} contains scatter plots in selected features of all main clusters.
For each main cluster, the scatter plot between two features has been selected that show a clear separation between the subclusters.

These scatter plots are used to assign meaningful names to the subclusters.
As already stated in the main text, it should be realized that the indicators \emph{high} and \emph{low} are relative to their main clusters.

\begin{figure}[h!]
    \centering
    \includegraphics[width=\linewidth]{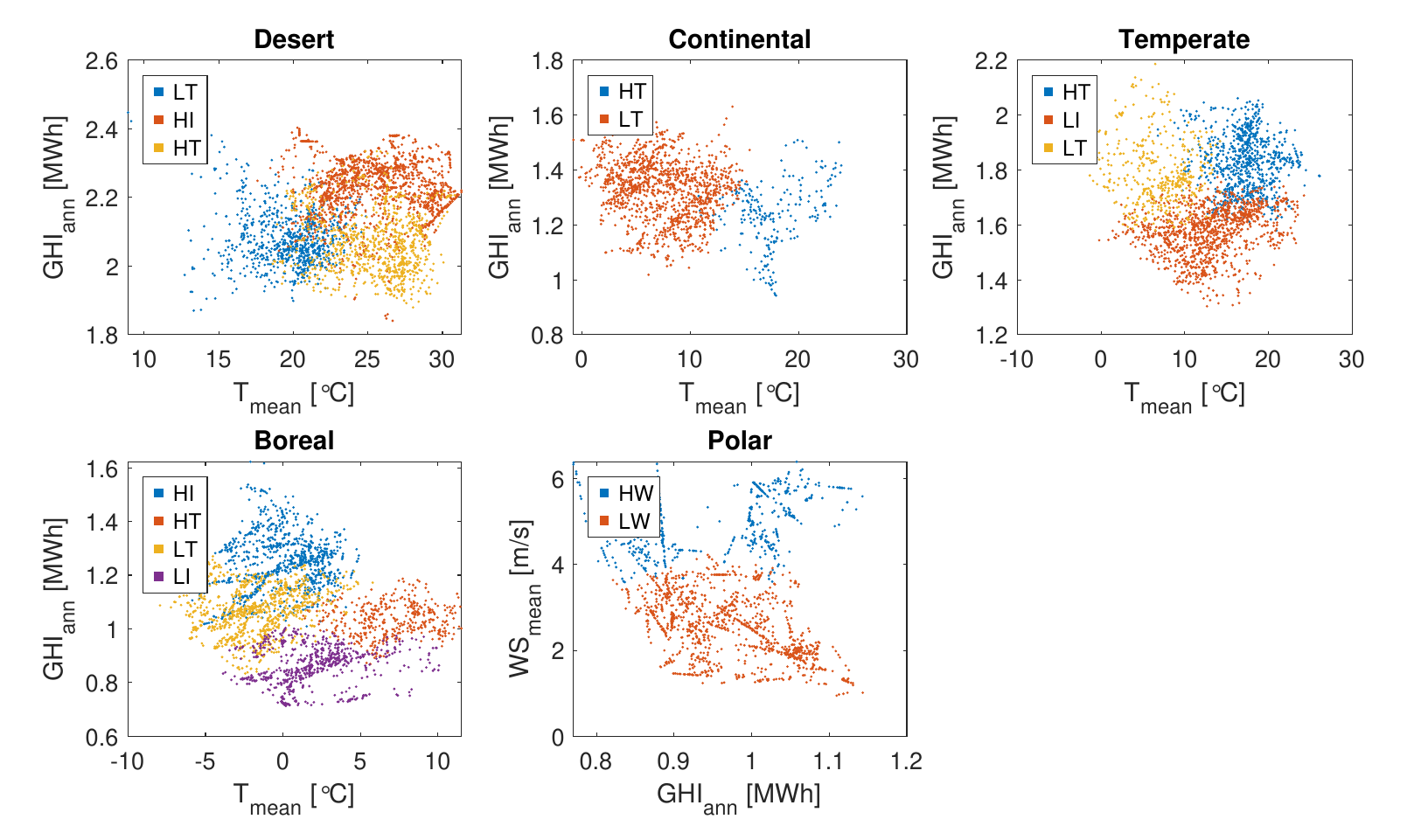}
    \caption{Scatter plots of the relevant features of the different subclusters in each main cluster. This is used to create meaningful annotations for the subclusters. Please note that different colors have been selected to more clearly indicate the different subclusters.}
    \label{fig:explainLabeling}
\end{figure}

\section{Overlap with existing classifications}
\label{App:OverlapExisting}
\setcounter{figure}{0} 
To investigate the similarity between the created climate classification and the already existing ones, Figure~\ref{fig:OverlapClassifications} presents the overlap the map in Figure~\ref{fig:ClimateClassification} and the one generated by Ascencio-Vásquez~\textit{et al.}~\cite{Ascencio-Vasquez2019MethodologyPerformance} and Triana de las Heras~\textit{et al.}~\cite{TrianadelasHeras2025AClassification}.
It can be seen that is a significant overlap between the climate zones, showing a  >70\% overlap for multiple clusters.
Still, there are also significant differences, highlighting the need for the classification developed in this work.

\begin{figure}[h!]
    \centering
    \includegraphics[width=\linewidth]{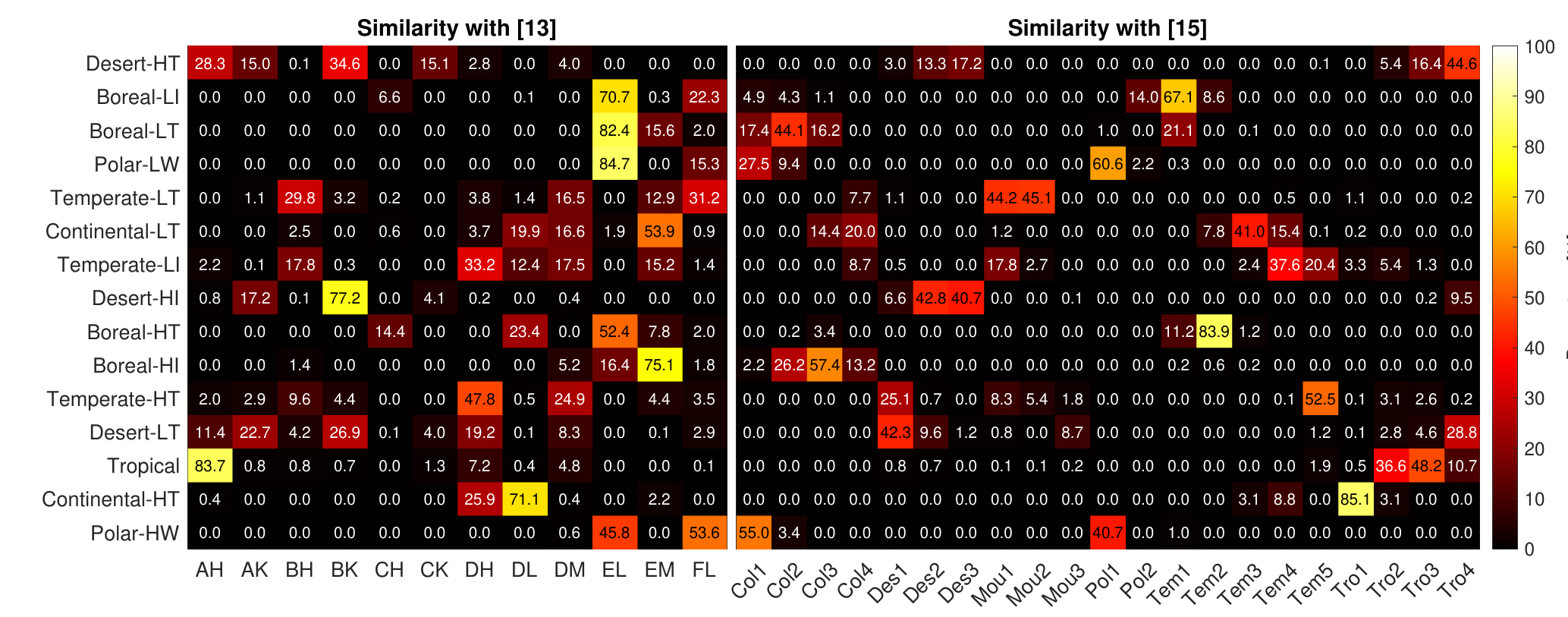}
    \caption{The overlap between the generated classification and the already existing ones. The percentages indicate which fraction of the climate zone from this work is also in the climate zone from \cite{Ascencio-Vasquez2019MethodologyPerformance} or \cite{TrianadelasHeras2025AClassification}. Therefore, the values in all rows equal to 100\%.}
    \label{fig:OverlapClassifications}
\end{figure}

\section{Definition discounted lifetime energy yield}
\label{App:DefinitionDisLEY}
\setcounter{figure}{0}
Figure~\ref{fig:BoxPlotsLEY} in the main text shows the boxplot of the discounted lifetime energy yield ($LEY_{dis}$) for all climate zones.
$LEY_{dis}$ for each location is calculated with
\begin{equation}
    LEY_{dis} = \sum_{t=1}^{LT} \frac{EY \cdot \left(1-0.2\cdot \frac{t}{LT}\right)}{(1+r_d)^t} ,
\end{equation}
where $LT$ is the module lifetime, $EY$ is the annual energy yield, $r_d$ is the discount rate (taken as 7\%~\cite{IEA2020_ProjectedElectricity}), and $t$ represents the years from 1 till $LT$.

The term $1-0.2\frac{t}{LT}$ is used to account for a degradation-induced reduction in energy yield throughout the lifetime of the module.
It assumes a linear degradation rate, needed to obtain an 80\% module performance at the end of the module lifetime.

\section{Representative locations for all climate zones}
\label{App:RepresentativeLocations}
\setcounter{table}{0}
To enable consistent global comparisons, a single representative location is identified for each climate zone.
For a given climate zone $k$, the mean weighted value of feature $j$, denoted by $\bar{x}_j$, is computed as

\begin{equation}
    \bar{x}_j = \frac{\sum_{i\in k} x_{i,j}\cdot S_j}{N_{k}},
\end{equation}
where $x_{i,j}$ represents the value of feature $j$ at location $i$, $S_j$ is the feature importance score, and $N_k$ denotes the total number of locations within climate zone $k$.
The representative location for each climate zone is then selected as the location whose feature values are closest to the corresponding mean weighted feature values.
Table~\ref{tab:Representativelocations} summarizes the geographic coordinates of the representative locations identified for each climate zone.

\begin{table}[h]
    \centering
    \caption{The coordinates of the representative locations for all climate zones.}
    \begin{tabular}{c|c|c|c}
        \textbf{Climate zone} & \textbf{Latitude} & \textbf{Longitude}& \textbf{Country}  \\
        \hline
        Desert-LT&	-19&	30& Zimbabwe \\
        Desert-HI&	-21&	129& Australia \\
        Desert-HT&	8&	42 & Ethiopia\\
        Tropical&	22&	95& Myanmar \\
        Continental-HT&	29&	117& China \\
        Continental-LT&	45&	-89& United States of America \\
        Temperate-HT&	-32& 27&  South Africa\\
        Temperate-LI&	39&	32&  Türkiye\\
        Temperate-LT&	39&	101& China \\
        Boreal-HI&	54&	-98& Canada \\
        Boreal-HT&	54&	33& Russia \\
        Boreal-LT&	59&	100& Russia \\
        Boreal-LI&	62&	9& Norway \\
        Polar-HW&	67&	-109& Canada \\
        Polar-LW&	67&	166& Russia \\
    \end{tabular}
    \label{tab:Representativelocations}
\end{table}

\bibliographystyle{elsarticle-num} 
\bibliography{References}

\end{document}